\documentclass[12pt]{article}

\usepackage[english]{babel}
\usepackage{graphicx,rotating,a4}

\def\ifmath#1{\relax\ifmmode #1\else $#1$\fi}%
\def\TeV{\ifmmode {\,\mathrm{ Te\kern -0.1em V}}\else
                   \textrm{Te\kern -0.1em V}\fi}%
\def\GeV{\ifmmode {\,\mathrm{ Ge\kern -0.1em V}}\else
                   \textrm{Ge\kern -0.1em V}\fi}%
\def\MeV{\ifmmode {\,\mathrm{ Me\kern -0.1em V}}\else
                   \textrm{Me\kern -0.1em V}\fi}%
\def\keV{\ifmmode {\,\mathrm{ ke\kern -0.1em V}}\else
                   \textrm{ke\kern -0.1em V}\fi}%
\def\eV{\ifmmode  {\,\mathrm{ e\kern -0.1em V}}\else
                   \textrm{e\kern -0.1em V}\fi}%

\newcommand{\fbi}{\,\rm{fb}^{-1}}

\newcommand{\ee}    {\mathrm{e}^+\mathrm{e}^-}

\newcommand{\ALR}    {A_{\mathrm{LR}}}

\newcommand {\Rb}   {\ifmath{R_{\mathrm{b}}}}

\newcommand{\MH}      {m_{\mathrm{H}}}

\newcommand{\MZ}      {m_{\mathrm{Z}}}
\newcommand{\MW}      {m_{\mathrm{W}}}
\newcommand{\MT}      {m_{\mathrm{t}}}
\newcommand{\GZ}      {\Gamma_{\mathrm{Z}}}

\newcommand{\Ghad}       {\Gamma_{\mathrm{had}}}

\newcommand {\so}   {\sigma_0^{\rm{had}}}
\newcommand {\stl}  {\sin^2 \theta_{\rm{eff}}^\ell}

\newcommand {\cAe} {\mbox{$\cal A_{\rm e}$}}
\newcommand {\cAb} {\mbox{$\cal A_{\rm b}$}}

\newcommand{\swsqeffl}    {\sin^2\!\theta_{\rm{eff}}^\ell}

\newcommand{\ppl}  {{\cal P}_{\rm{e}^+}}
\newcommand{\pmi}  {{\cal P}_{\rm{e}^-}}
\newcommand{\ppm}  {{\cal P}_{\rm{e}^\pm}}
\newcommand{\peff}  {{\cal P}_{\rm{eff}}}
\begin{document}
%################################################## titlepage declaration
\begin{titlepage}

\pagenumbering{arabic}
\begin{flushright}
%LC-PHSM-2000-??-TESLA\\
{\today}
\end{flushright}
%========================================================================%
\vspace*{2.cm}
\begin{center}
\Large 
\boldmath
{\bf
%===================> DELPHI note title        =====> To be filled <=====%
What is the Case for a Return to the Z-Pole?
%========================================================================%
} \\
\unboldmath
\vspace*{2.cm}
\normalsize { 
%===================> DELPHI note author list  =====> To be filled <=====%
   {\bf K. M\"onig}\\
   {\footnotesize DESY-Zeuthen}\\
   
%========================================================================%
}
%\vspace*{2.cm}
\end{center}
\vspace{\fill}
\begin{abstract}
\noindent
The possibilities to run with a linear collider at the Z-pole with
high luminosity are examined.
Apart from the implications on machine and detector the interest for
electroweak and B-physics is discussed.
\end{abstract}
\vspace{\fill}
\begin{center}
%==========> Proceedings.. presented at ..==> To be filled if needed<=====%
Plenary talk at the International Workshop on Linear Colliders,  \\
Fermilab 24 - 28 October 2000
%=========================================================================%
\end{center}
\vspace{\fill}
\end{titlepage}
\section{Introduction}

In the past LEP and SLC have contributed a lot to our knowledge of
particle physics. On one hand the measurements of the Z-mass and couplings
near the Z pole and the measurement of the W mass above the W-pair
threshold test the Standard Model at the loop level and open a window to
new physics at much higher scales through virtual effects. On the other
hand the large cross section at the Z pole for all fermions apart from
t-quarks provides a rich source for some particles like $\tau$'s, D-mesons
or B-mesons that can be studies in a rather clean environment.

A linear collider can potentially repeat all the measurements done at LEP
and SLC with much higher statistics. It is therefor worth studying in
which fields Z-running at such a machine can still contribute in the light
of the competition from new machines, namely the $\ee$ B-factories, run II
of the {\sc Tevatron} and the LHC.

\section{The Giga-Z setup}

The Linear Collider in the so called Giga-Z scenario should be able to
provide about $10^9$ recorded Z-decays. As it will be discussed later this
is only useful if also a high degree of electron polarization ($\ge
80\%$) is possible. In addition one needs either a very good precision on
polarimetry (${\cal O}(0.1-0.5\%)$) or positron polarization of
at least 20\%.

In addition to Z running also running with similar luminosity at the
W-pair threshold should be possible.

In the NLC design \cite{ref:nlc} the positron source under study is
independent of the electron beam. It is thus feasible to start the machine
around the Z-pole and upgrade it to higher energies later.

The TESLA design \cite{ref:tesla} uses the high energy ($>150 \GeV$)
electron beam to produce the positrons by sending the beam through a
wiggler or a helical undulator. To run at the Z-pole one needs one part of
the electron arm to produce the $45 \GeV$ physics beam and the other
part to accelerate the high energy beam for positron creation. In this
design one would therefor prefer to come back to the Z pole only after the
basic high energy program is completed.

Table \ref{tab:lum} shows the luminosity, beamstrahlung and depolarization
for the two designs. The option with low beamstrahlung is shown for the NLC
as an example, however it is possible with all designs.

\begin{table}[htb]
\begin{center}
  \begin{tabular}[c]{|l|c|c|c|}
    \hline
      & \multicolumn{2}{|c|}{NLC} & TESLA \\
      \cline{2-3}
      & norm & low $\delta_B$ & \\
    \hline
    ${\cal L} \, (10^{33})$ & 4.1 & 2  & 5 \\
    $\delta_B \, (\%)$ & 0.16 & 0.05 & 0.1 \\
    $\Delta {\cal P}_{\rm{IP}}\, (\%)$ & 0.07& 0.02 & 0.1 \\
    \hline
  \end{tabular} 
\end{center} 
\caption{ Luminosity, beam\-strah\-lung and
de\-po\-la\-ri\-sation in the interaction point for the two designs. Beamstrahlung
and depolarization are given for the outgoing beam. For the interacting
particles they are a factor two to four smaller. } \label{tab:lum}
\end{table}
 
The total hadronic cross section at the Z pole is given by $\sigma \approx
\sigma_u (1 + \ppl \pmi)$ with $\ppl$ ($\pmi$) being the positron
(electron) polarization and $\sigma_u \approx 30 \rm{nb}$. $10^9$ Zs can
thus be be produced in 50-100 days of running with a Z rate of 100-200Hz.
If it is found worthwhile also $10^{10}$ Zs can be produced in three to
five years (150days/year) of running time.

The beamstrahl effects seem manageable and the depolarization in the
interaction is almost negligible.

\section{Electroweak Physics}

The interesting quantities in electroweak physics, accessible at Giga-Z
are: 
\begin{itemize} 
\item the normalization of the Z axial-vector
 coupling to leptons ($\Delta \rho_\ell$) which is measured from the
 partial width $\Gamma_\ell$; 
\item the effective weak mixing angle measured
 from the ratio of vector to axial vector coupling of $Z \rightarrow \ell \ell$
 ($\stl$); 
\item the mass of the W ($\MW$);
\item the strong coupling
 constant from the Z hadronic decay rate ($\alpha_s (\MZ^2)$);
\item the vertex correction to Zbb vertex measured from $\Rb, \cAb$. 
\end{itemize}
From the parameters listed above $\Delta \rho_\ell$ and $\alpha_s$ are
obtained from a scan of the Z-resonance. Table \ref{tab:lepscan} shows the
LEP precision on the minimally correlated observables.
$\alpha_s$ can be obtained from $R_\ell$ only. For $\rho_\ell$,
however, one needs to improve on all parameters apart from $\MZ$, so
that a scan and an absolute luminosity measurement are needed.

\begin{table}
\begin{center}
  \begin{tabular}[c]{|l|c|}
\hline
           & LEP precision\\
\hline
$\MZ$      & $0.2 \cdot 10^{-4}$ \\
$\GZ$      & $0.9 \cdot 10^{-3}$ \\
$\so = \frac{12 \pi}{\MZ^2} \frac{\Gamma_e \Ghad}{\Gamma_Z^2} $
                    & $0.9 \cdot 10^{-3}$ \\
$R_\ell = \frac{\Ghad}{\Gamma_l} $ 
                    & $1.2 \cdot 10^{-3}$ \\
\hline
  \end{tabular}
\end{center}
\caption[]{LEP precision on the minimally correlated ob\-ser\-vables from the 
  Z-scan \cite{ref:lepres}.
  }\label{tab:lepscan} 
\end{table}

A measurement of the beam energy relative to the Z-mass of $10^{-5}$
seems possible. This would improve the relative accuracy on $\GZ$ to 
$0.4 \cdot 10^{-3}$. The selection efficiency for muons, taus and
hadrons should be improved by a factor three relative to the best LEP
experiment \cite{ref:marc}, resulting in 
$\Delta R_\ell / R_\ell = 0.3 \cdot 10^{-3}$.
Also the experimental error on the luminosity might improve by this
factor. However if the theoretical uncertainty stays at its present value
(0.05\%) the possible precision on $\so$ is only $0.6 \cdot 10^{-3}$.
It should however be noted that beamstrahlung plus energy spread
increase the fitted $\GZ$ by about $60 \MeV$ and decrease $\so$ by 1.8\%,
where the majority comes from the beamspread.
For the $\alpha_s$ measurement these effects need to be understood to
roughly 10\% while for $\Delta \rho_\ell$ one needs 2\%. There is the
potential to achieve this precision with the acolinearity measurement
of Bhabha events \cite{ref:bsmeas} or to extend the scan to five scan
points and fit for the beamspread, but both options need further studies.

Table \ref{tab:scanres} compares the parameters that can be obtained
from a scan at Giga-Z with the results obtained at LEP. Gains of a
factor two to three are generally possible.
\begin{table}
\begin{center}
\begin{tabular}[c]{|c|c|c|}
\hline
 & LEP\cite{ref:lepres} & Giga-Z \\
\hline
$\MZ$ & {$ 91.1874 \pm 0.0021 \GeV$} & {$ \pm 0.0021 \GeV$} \\
$\alpha_s(\MZ^2)$ & {$ 0.1183 \pm 0.0027 $} & {$ \pm 0.0009$} \\
$\Delta \rho_\ell$ & {$ (0.55 \pm 0.10 ) \cdot 10^{-2}$} 
& {$ \pm 0.05\cdot 10^{-2}$}  \\
$N_\nu$ & {$ 2.984 \pm 0.008 $} & {$ \pm 0.004 $} \\
\hline
\end{tabular}
\end{center}
\caption{Comparison between Giga-Z and LEP for results obtained from a Z-scan.
  }
\label{tab:scanres} 
\end{table}

Much more interesting are the prospects on $\stl$ \cite{ref:epj,ref:peter}. 
With polarized beams the most sensitive observable is the left-right
cross section asymmetry $\ALR$:
\begin{eqnarray}
\label{eq:alrdef}
\ALR & = & \frac{1}{{\cal P}}\frac{\sigma_L-\sigma_R}{\sigma_L+\sigma_R}\\
     & = &   \cAe \nonumber \\
     & = & \frac{2 v_e a_e}{v_e^2 +a_e^2} \nonumber \\
  {v_e}/{a_e} & = & 1 - 4 \stl \nonumber
\end{eqnarray}
independent of the final state.
With $10^9$ Zs, an electron polarization of 80\% and no positron
polarization the statistical error is $\Delta \ALR = 4 \cdot 10^{-5}$.
The error from the polarization measurement is
$\Delta \ALR/\ALR = \Delta {\cal P}/{\cal P}$.
With electron polarization only and 
$\Delta {\cal P}/{\cal P} = 0.5\%$  one has
$\Delta \ALR = 8 \cdot 10^{-4}$. 
If also positron polarization is available ${\cal P}$ in equation
(\ref{eq:alrdef}) has to be replaced by 
$\peff \, = \, \frac{\ppl+\pmi}{1+\ppl\pmi}$. 
For $\pmi(\ppl) = 80\%(60\%)$, due to error propagation, the
error in $\peff$ is a factor of three to four smaller than the error on
$\ppl,\, \pmi$ depending on the correlation between the two
measurements \cite{ref:fujii}.

\begin{figure}[htbp]
  \begin{center}
    \includegraphics[height=6cm,bb=0 6 567 472]{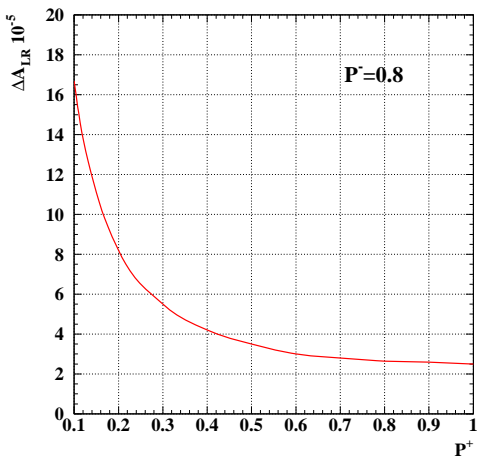}
  \end{center}
  \caption{Error of $\ALR$ as a function of the positron
    polarization for a luminosity corresponding to $10^9$ unpolarized Zs.}
  \label{fig:polerr}
\end{figure}

However, with positron polarization a much more precise measurement is
possible using the Blondel scheme \cite{ref:alain}. 
The total cross section with both beams being polarized is given as
  $\sigma \, = \, \sigma_u \left[ 1 - \ppl \pmi + \ALR (\ppl - \pmi) \right]$.
If all four helicity combinations are measured $\ALR$ can be determined
without polarization measurement as
\[
\ALR \, = \, \sqrt{\frac{
    ( \sigma_{++}+\sigma_{-+}-\sigma_{+-}-\sigma_{--})
    (-\sigma_{++}+\sigma_{-+}-\sigma_{+-}+\sigma_{--})}{
    ( \sigma_{++}+\sigma_{-+}+\sigma_{+-}+\sigma_{--})
    (-\sigma_{++}+\sigma_{-+}+\sigma_{+-}-\sigma_{--})}}
\]
Figure \ref{fig:polerr} shows the error on $\ALR$ as a function on the
positron polarization.  For $\ppl > 50\%$ the dependence is relatively weak.
For $10^9$ Zs a positron polarization of 20\% is better than a
polarization measurement of $0.1\%$ and electron polarization only.

However polarimeters for relative measurements are still needed. The
crucial point is the difference between the absolute values of the
left- and the right-handed states.
If the two helicity states for electrons and positrons are written as 
$\ppm = \pm |\ppm| + \delta \ppm$ the
dependence is $\rm{d} \ALR / \rm{d} \delta \ppm \approx 0.5$.
One therefor needs to understand $\delta \ppm $ to $< 10^{-4}$.
To achieve this a polarimeter with at least two channels
with different analyzing power to handle effects like polarization
dependent e-$\gamma$ luminosity internally is needed.

Due to $\gamma-Z$-interference the dependence of $\ALR$ on the beam
energy is $\rm{d} \ALR / \rm{d} \sqrt{s} = 2 \cdot 10^{-2}/\GeV$. The
difference $\sqrt{s} - \MZ$ thus needs to be known to $\sim 1 \MeV$.
For the same reason beamstrahlung shifts $\ALR$ by $\sim 9 \cdot 10^{-4}$
(TESLA design), so its uncertainty can only be a few percent.
If beamstrahlung is identical in the Z-scan to calibrate the
beam energy it gets absorbed in the energy calibration, so that
practically no corrections are needed for $\ALR$.
How far the beam parameters can be kept constant during the scan and
how well the beamstrahlung can be measured still needs further studies.
However, for $\ALR$ only the beamstrahlung and not the beamspread matters.
If the beamstrahlung cannot be understood to the required level in the
normal running mode one can still go to a mode with lower beamstrahlung
increasing the statistical error or the running time.

Other systematics should be small. A total error of $\Delta \ALR =
10^{-4}$ will thus be assumed,
corresponding to $\Delta \swsqeffl = 0.000013$. This is an improvement
of a factor 13 relative to the combined LEP/SLD \cite{ref:lepres} result.

For the b-quark observables one profits from the improved b-tagging
and, in the case of the forward backward asymmetries, also from beam
polarization. In summary for $\Rb$ a factor five improvement and for
$\cAb$ a factor 15 is possible relative to LEP/SLD \cite{ref:epj,ref:bruce}.
If the present slight discrepancy between the measured and the
predicted $\cAb$ is real it cannot be missed with Giga-Z.

The cleanest way to measure the W-mass is from a threshold scan where
no uncertainties from fragmentation, color reconnection etc. enter.
Near threshold the cross section is dominated by the neutrino
t-channel exchange with a phase space suppression of $\beta$, compared
to the $\beta^3$ suppressed s-channel. The hard process is thus
dominated by the well understood We$\nu$-coupling.
However, for the radiative corrections a full one loop calculation is
needed, since the double pole approximation \cite{ref:wwloop} is not
applicable in this region.
To estimate the obtainable precision on $\MW$ a scan around 
$\sqrt{s}=161 \GeV$ has been simulated \cite{ref:graham} with
${\cal L} = 100 \fbi$, corresponding to one year of running.
Beam polarization has been used to enlarge the signal and to measure
the background which requires a polarization measurement of
$\Delta {\cal P}/{\cal P} < 0.25\%$.
Assuming efficiencies and purities as at LEP the error on $\MW$ will be
$6 \MeV$ if the luminosity and the efficiencies are known to $0.25
\%$, increasing to $7 \MeV$ if they are left free in the fit.

Before the precision data can be interpreted in the framework of the
Standard Model or one of its extensions the uncertainties of the
predictions stemming from the uncertainties in the input parameters
need to be discussed.
At the moment the by far largest uncertainty, especially for the
prediction of $\stl$ comes from the running of the electromagnetic
coupling from zero to the Z-scale.
Using data only, ignoring the latest BES results, the uncertainties on
$\stl$ and $\MW$ are $\Delta \swsqeffl = 0.00023$, 
$\Delta \MW=12\MeV$ \cite{ref:aem_data}. Using perturbative QCD also
at lower energies, these errors can be reduced by about a factor of
three \cite{alpha_d,alpha_k}. 
Only if the hadronic cross section up to the $\Upsilon$ is known to
1\% the errors can be brought down to
$\Delta \swsqeffl = 0.000017$, $\Delta \MW<1\MeV$ \cite{ref:aem_j}.

The $2 \MeV$ error on the Z mass induces an error of 0.000014 on
$\stl$ and, if the beam energy is calibrated relative to the Z-mass
$1 \MeV$ on $\MW$.
Unless a new circular collider for Z-pole running is built, where it
is easier to measure the absolute energy scale, this error limits
even further improvement on $\swsqeffl$.

A $1 \GeV$ error on $\MT$ gives $\Delta \swsqeffl = 0.00003$,
$\Delta \MW=6\MeV$. With an error of $\Delta \MT \approx 100 \MeV$ as
it is expected from a threshold scan at a linear collider, the
contribution from $\MT$ on the predictions will be negligible.

\begin{figure}[htb]
  \begin{center}
    \includegraphics[height=6cm,bb=0 0 567 473]{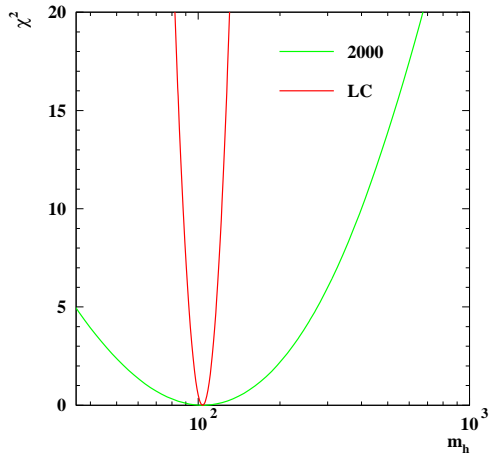}
  \end{center}
  \caption{$\Delta \chi^2$ as a function of $\MH$ for the present and
    the Giga-Z data.
    }
  \label{fig:higgschi} 
\end{figure}

If it is assumed that the Standard Model is the final theory the Higgs
mass can be determined from the Giga-Z data to a 
precision of 5\%\cite{ref:epj,ref:sven}.
Figure \ref{fig:higgschi} compares the sensitivity of the Higgs fit to
the Giga-Z data with the present situation \cite{ref:lepres}.

\begin{figure}[htb]
  \begin{center}
    \includegraphics[height=6cm]{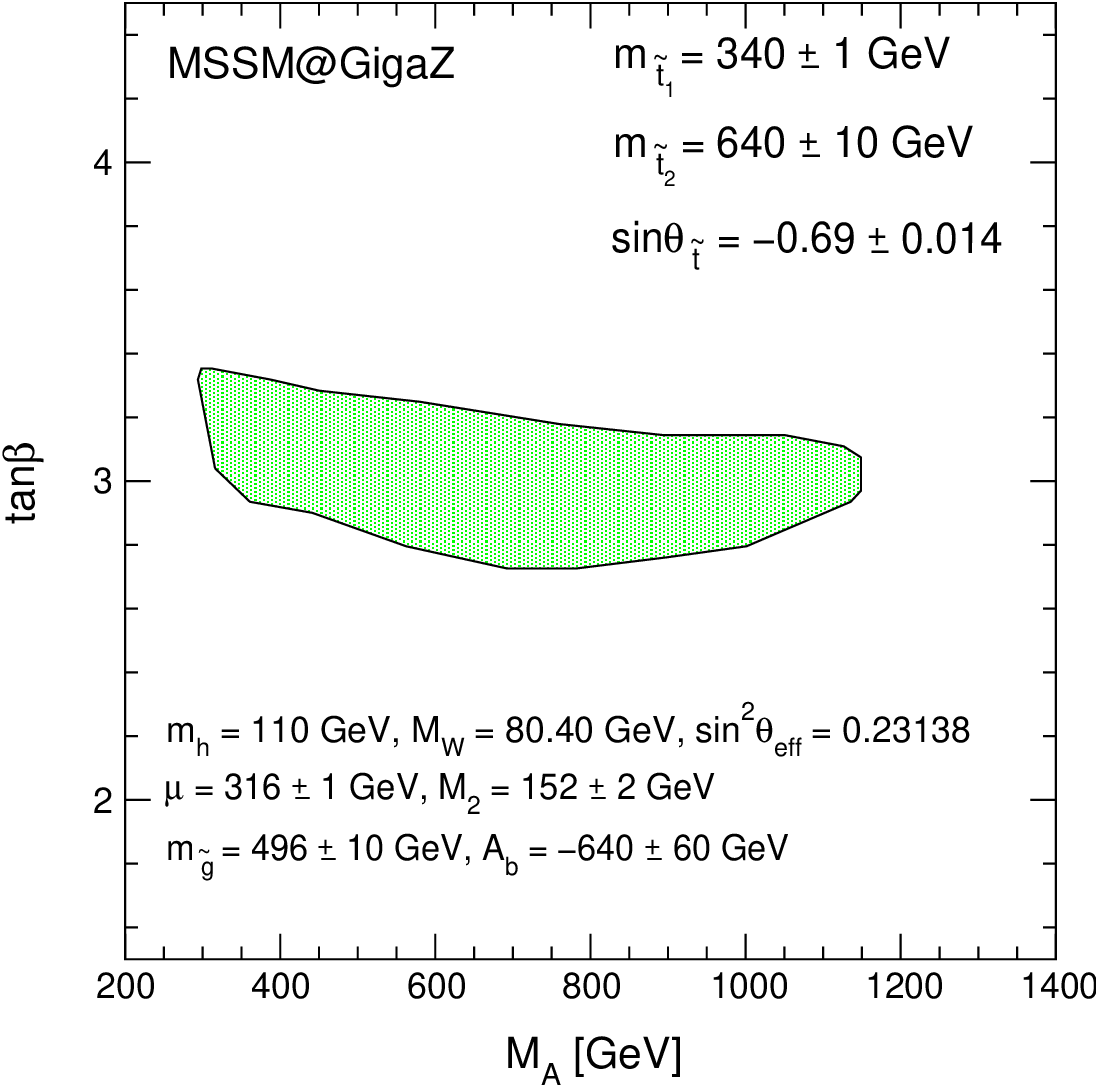}
  \end{center}
  \caption{ The region in the $m_{\rm{A}}- \tan \beta$ plane, allowed by
    $1\,\sigma$ errors by the measurements of $\MW$ and $\swsqeffl$.
    }
  \label{fig:mssm}
\end{figure}

Also in extensions of the Standard Model the precision data can be
used to predict model parameters. As an example figure \ref{fig:mssm}
shows a possible prediction for $m_{\rm{A}}$ and $\tan \beta$ in the
MSSM, once the light Higgs is found and the parameters of the stop
sector are measured \cite{ref:sven}.

\begin{figure}[p]
  \begin{center}
    \includegraphics[height=6.cm,bb=0 0 567 506]{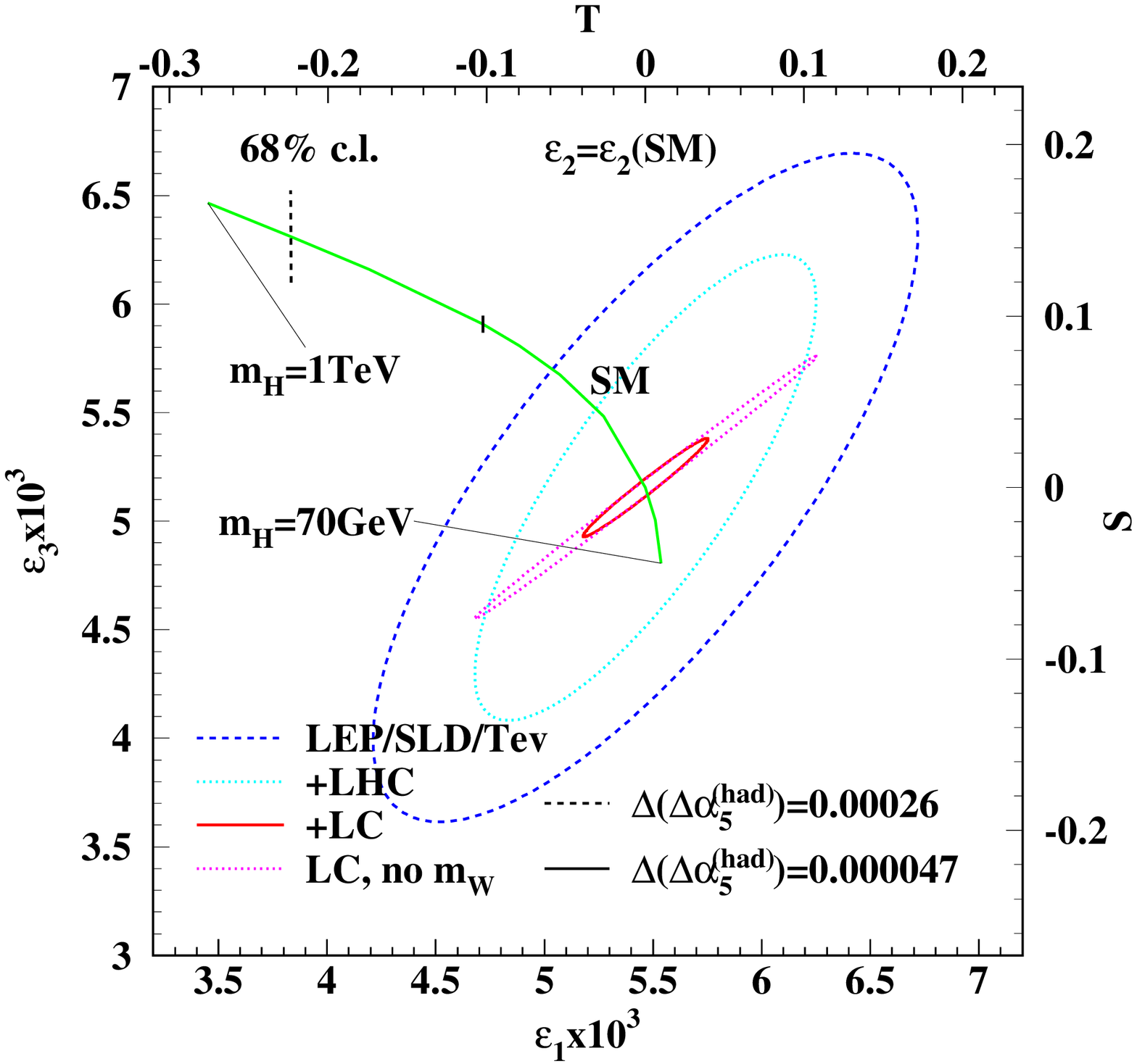}
  \end{center}
 \vspace{-0.25cm}
  \caption{Allowed regions in the $\varepsilon_1 - \varepsilon_3$ (S-T) planes
    for various assumptions compared to the Standard Model prediction.
    Also shown is the uncertainty in the prediction due to $\alpha(\MZ)$.
    }
  \label{fig:eps} 
%\vspace{-0.5cm}
\end{figure}

The data can also be analyzed within the model independent 
$\varepsilon$ \cite{ref:eps} or STU \cite{ref:stu} parameters.
Figure \ref{fig:eps} shows the allowed regions in the 
$\varepsilon_1 - \varepsilon_3$ (S-T) plane with $\varepsilon_2$ (U)
fixed to its Standard Model prediction for various configurations
compared to the Standard Model. Removing $\MW$ from the fit is
equivalent to not constraining $\varepsilon_2$ (U). The tight
constraint in the direction of the SM trajectory is dominated by the
$\stl$ measurement, while the orthogonal direction mainly profits from $\MW$.

\begin{figure}[p]
  \begin{center}
    \includegraphics[height=11.cm,bb=3 23 510 531]{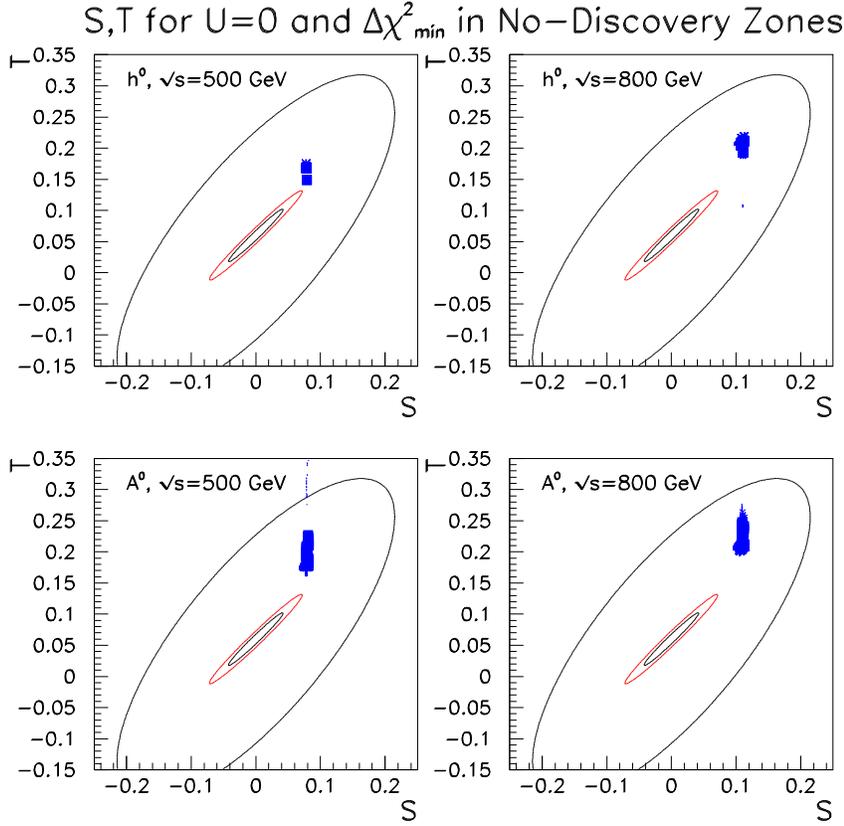}
  \end{center}
\vspace{-0.25cm}
\caption[bla]{Prediction for S and T from the 2 Higgs doublet model with a
  light Higgs for the cases where no Higgs is found compared to the
  current electroweak data (95\% c.l.) and the projection for 
  Giga-Z (95\% and 99\% c.l.).
  }
  \label{fig:thdm} 
\end{figure}

Figure \ref{fig:thdm} shows as an application the S,T-predictions from
the 2-Higgs doublet model (2HDM) for the cases, where a light Higgs exists,
but cannot be seen \cite{ref:gunion} compared to present and Giga-Z data.
Only Giga-Z allows to distinguish the Standard Model with a light
Higgs from the 2HDM, but this also needs the precise measurement of $\MW$.

\section{B-physics}

Within $10^9$ Z-decays about $4 \cdot 10^8$ B-hadrons are
produced. This is a sample comparable to the
$\ee$-B-factories. Compared to these machines the large boost allows a
good separation of the two B's in the event and gives a much better
decay length resolution. 
In addition all B-hadron species are produced, while at the $\Upsilon (4S)$
only $B^0$ and $B^\pm$ are present.

Compared to the experiments at hadron colliders (BTeV, LHCb) the
statistics is much smaller. However all events are triggered and the
environment is much cleaner.

Due to the large forward backward asymmetry
with polarized beams one gets also a very efficient tagging of the
initial state flavor from the direction of the event axis only.
In all other machines the flavor must be tagged by reconstructing the
other B in the event. 

Up to now the studies mainly repeat the ones done for the other
machines. Only very little exists on reactions that cannot be
measured somewhere else.

\begin{figure}[htbp]
  \begin{center}
    \includegraphics[height=6.cm]{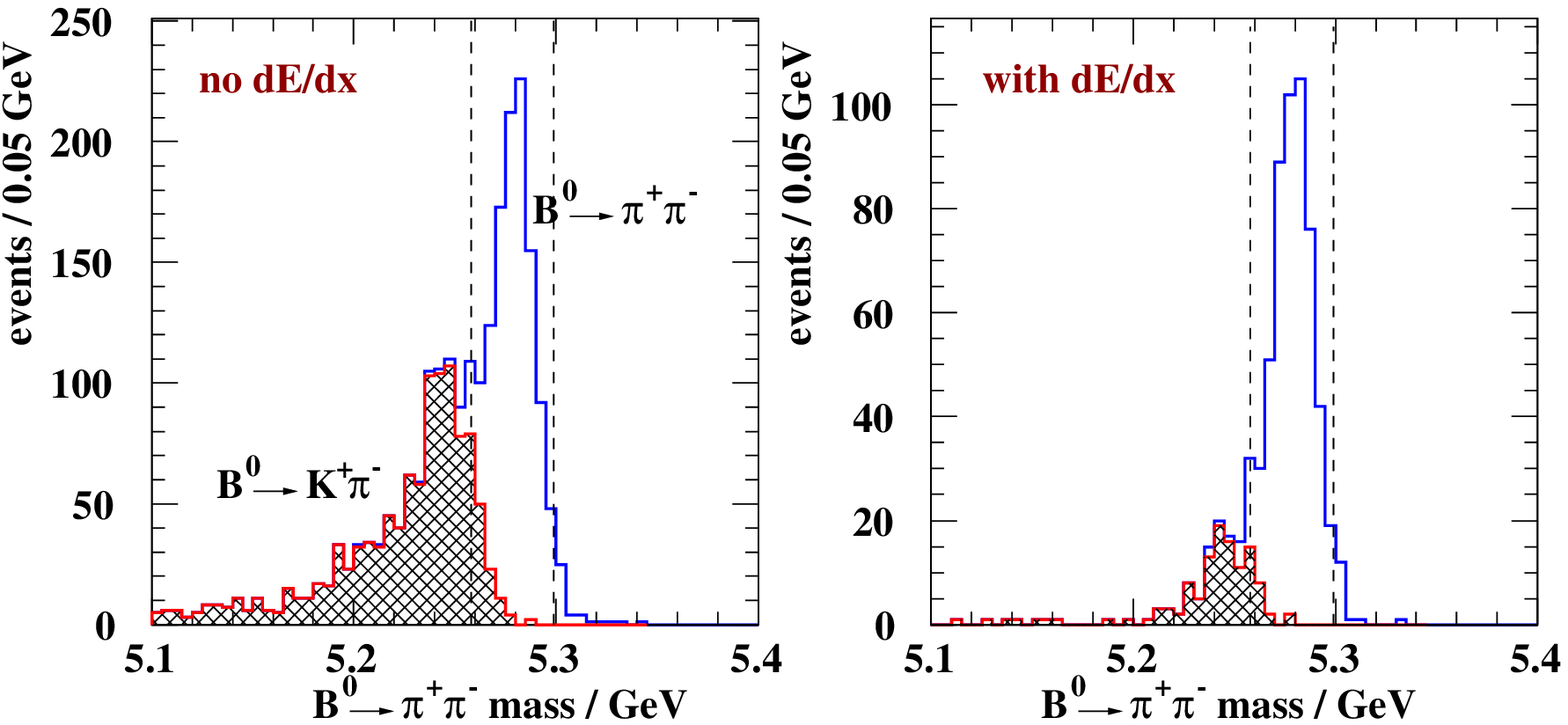}
  \end{center}
  \caption{$\pi^+ \pi^-$ mass resolution for $B^0 \rightarrow \pi^+ \pi^-$ 
    and $B^0 \rightarrow K^+ \pi^-$ without and with 
    a cut on dE/dx in the TPC.
    }
  \label{fig:bpipi}
\end{figure}

To understand CP-violation in B-decays the time dependent asymmetries in
$B^0 \rightarrow J/\Psi K^0_s$ have been analyzed to measure 
$\sin 2 \beta$ and in $B^0 \rightarrow \pi^+ \pi^-$ to measure 
$\sin 2 \alpha$ \cite{ref:epj}. 
The analysis of $B^0 \rightarrow J/\Psi K^0_s$ is
experimentally relatively easy. For $B^0 \rightarrow \pi^+ \pi^-$ 
one needs to separate the mode from $B^0 \rightarrow K^+ \pi^-$.
Figure \ref{fig:bpipi} shows the reconstructed invariant mass for
these two modes under the $\pi^+ \pi^-$ hypotheses for the TESLA
detector. The excellent mass resolution separates the two modes already
very well and the remaining background can be rejected using dE/dx in
the TPC.

Table \ref{tab:cpres} compares the Giga-Z reach with other
machines for $\sin 2 \alpha$ and $\sin 2 \beta$ under the assumption that 
the penguin contributions to $B^0 \rightarrow \pi^+ \pi^-$ are negligible.
For $10^9$ Zs interesting cross checks are possible while for
$10^{10}$ Zs the linear collider gets highly competitive.

\begin{table}[htb]
  \begin{center}
    \begin{tabular}[c]{|l|c|c|}
      \hline
      & $\sin 2 \beta$ & ``$\sin 2 \alpha$'' \\
      \hline
      BaBar/Belle & $0.12$ & $0.26$ \\
      CDF    & $0.08$ & $0.10$ \\
      BTeV/year & $0.03$ & $0.02$ \\
      ATLAS  & $0.02$ & $0.14$ \\
      LHC-b  & $0.01$ & $0.05$ \\
      \hline
      Giga-Z ($10^9$ Zs) & $0.04$ & $0.07$ \\
      \hline
    \end{tabular}
  \end{center}
  \caption{Precision on the angles of the unitarity triangle for the
      different machines.
      } 
    \label{tab:cpres} 
\end{table}

To disentangle the penguin contributions to $B^0 \rightarrow \pi^+ \pi^-$ the
branching ratios $\rm{BR}(B^0 \rightarrow \pi^0 \pi^0)$ and
$\rm{BR}(B^+ \rightarrow \pi^+ \pi^0)$ can be measured. With $10^9$ Zs
the precision at Giga-Z is similar to the $\ee$-B-factories.

The large boost and the good vertex resolution also offers the
possibility to measure $B_s \overline{B_s}$-mixing \cite{ref:epj,ref:bruce}.
In the fully reconstructed mode $B_s \rightarrow D_s \pi, \, D_s \rightarrow
\Phi \pi, KK$ the detector resolution is about $40 \rm{ps}^{-1}$.
Probably $B_s \overline{B_s}$-oscillations will be discovered at that time,
but the good resolution should allow for a precise measurement of the
frequency. 

There are also some studies which are only possible in the Giga-Z
environment. As two examples \cite{ref:mannel} one can measure the branching 
ratio $\rm{BR}(B \rightarrow X_s \nu \bar{\nu})$ or one can test quark hadron
duality in B-decays, for example by measuring $V_{cb}$ in $B_s$-decays.
This assumption is essentially untested but is needed for the
interpretation of many results.

\section{Other physics topics}

In principle all LEP/SLD analyses can be repeated at Giga-Z. However
many of them are already now systematics limited or can be done
better at other machines. Not many detailed studies exist beyond the
ones already reported.

One field specific to $\ee$ colliders running on the Z pole is the
study of flavor violating Z decays. The lepton flavor violating decays
$Z \rightarrow e \tau, \mu \tau$ have been studied in detail \cite{ref:lpv}.
With $10^9$ events Giga-Z is sensitive to the $10^{-8}$ level. With
this precision one would be sensitive to the predictions of models
with heavy neutrinos in the $\TeV$ mass range or some supersymmetric models.

\section{Detector and machine issues}

In the present design NLC plans to run for Giga-Z with 180 bunch
trains per second, 190 bunches per train and a bunch spacing of 1.4ns.
On the contrary TESLA might run with 5 bunch trains per second, 
2800 bunches per train and a bunch spacing of 340ns.
In the NLC design the detector basically integrates over a full bunch
train, while for TESLA single bunches can be separated.
Table \ref{tab:zpertrain} shows the Z-multiplicity per NLC train or
TESLA bunch.

\begin{table}[htb]
\begin{center}
  \begin{tabular}[c]{|r|c|c|}
    \hline
     & NLC-train& TESLA-bunch \\
    \hline
    0 Zs            & $0.33$ & $0.986$       \\
    1 Z\phantom{s}  & $0.37$ & $1.4 \cdot 10^{-2}$ \\
    2 Zs            & $0.20$ & $9.7 \cdot 10^{-5}$ \\
    $\ge 3$ Zs      & $0.10$ & $4.5 \cdot 10^{-7}$ \\
    \hline
  \end{tabular}
\end{center}
\caption{Z-multiplicity in a NLC-train and TESLA bunch.
  } 
\label{tab:zpertrain} 
\end{table}
Z counting for the $\ALR$ measurement should be possible in both
cases. However how much a larger Z multiplicity inside an NLC train
affects more complicated B-physics analyses needs detailed studies.

The presently planned detectors provide excellent momentum resolution,
b-tagging and hermeticity and are more or less ideal for the
electroweak measurements.
The excellent vertexing also helps to separate different Zs within one
bunch or train using the different vertex positions along the beam axis.

For some analyses in B-physics a good
particle identification is required. Partly this can be replaced by
the superb invariant mass resolution and the large magnetic field helps
also to separate close-by particles to optimize the dE/dx resolution.
Past experience has shown that specialized particle identification
devices for this energy region tend to compromise the other features
of the detector
so that it is not clear in how far a detector optimized for B-physics
at Giga-Z would be different from the high energy one.

$10^9$ Zs can be recorded in 50 to 100 days of running. Since some Z
running is required for detector calibration in any case this can be
accommodated easily within the normal schedule of the machine.

If however $10^{10}$ Zs are found a worthwhile goal, which needs
several years of running, one might need a specialized interaction
region and detector that can run simultaneously with the high energy 
experiment.

\section{Conclusions}

With a relatively modest effort a huge gain in the precision
measurements on the Z-pole is possible. Also the W-mass can be
improved significantly if one year is spent for it. These measurements
allow stringent tests of the then-Standard-Model.
By no means the possibility to do these measurements should be
excluded by the machine and detector designs.

Some interesting cross check in B-physics can be done with $10^9$ Zs
and there might be the possibility to improve on the precision from
hadron machines with $10^{10}$ Zs. However this option needs further
studies and it is too early to conclude on that.

\section*{Acknowledgments}
I would like to thank the organizers of the conference for the superb
organization and the pleasant atmosphere at the meeting.
Also I'd like to thank J. Gunion, S. Heinemeyer, T. Raubenheimer,
P. Rowson, B. Schumm, N. Walker, G. Weiglein, and M. Woods for useful
discussions. 
%\pagebreak

\end{document}